# Syneruptive sequential fragmentation of pyroclasts from fractal modeling of grain size distributions of fall deposits: The Cretaio Tephra eruption (Ischia Island, Italy)

Joali Paredes[1]*, Daniele Morgavi[1], Mauro Di Vito[2], Sandro De Vita[2], Fabio Sansivero[2] Kai Dueffels[3], Gert Beckmann[3] & Diego Perugini[1]

[1] Department of Physics and Geology, University of Perugia, Piazza Università, Perugia 06123, Italy

[2] Istituto Nazionale di Geofisica e Vulcanologia, Osservatorio Vesuviano, Via Diocleziano 328, 80124 Napoli, Italy.

[3] Retsch Technology GmbH. Retsch Allee 1-5, 42781 Haan. Germany

*Corresponding author:*

Joali Paredes

e-mail: joali.paredes@studenti.unipg.it

Tel.: +39 075 585 2601

Fax: +39 075 585 2603

**Abstract**

In this work we used fractal statistics in order to decipher the mechanisms acting during explosive volcanic eruptions by studying the grain size distribution (GSD) of natural pyroclastic-fall deposits. The method was applied to lithic-rich proximal deposits from a stratigraphic section of the Cretaio Tephra eruption (Ischia Island, Italy). Analyses were performed separately on bulk material, juvenile, and lithic fraction from each pyroclastic layer. Results highlight that the bulk material is characterized by a single scaling regime whereas two scaling regimes, with contrasting power-law exponents, are observed for the juvenile and the lithic fractions. On the basis of these results, we infer that the bulk material cannot be considered as a good proxy for deducing eruption dynamics because it is the result of mixing of fragments belonging to the lithic and juvenile fraction, both of which underwent different events of fragmentation governed by different mechanisms. In addition, results from fractal analyses of the lithic fraction suggest that it likely experienced a fragmentation event in which the efficiency of fragmentation was larger for the coarser fragments relative to the finer ones. On the contrary, we interpret the different scaling regimes observed for the juvenile fraction as due to sequential events of fragmentation in the conduit, possibly enhanced by the presence of lithic fragments in the eruptive mixture. In particular, collisional events generated increasing amounts of finer particles modifying the original juvenile GSDs and determining the development of two scaling regimes in which the finer fragments record a higher efficiency of fragmentation relative to the coarser ones. We further suggest that in lithic-rich proximal fall deposits possible indications about the original GSDs of the juvenile fraction might still reside in the coarser particles fraction.

# 1. INTRODUCTION

The grain size distribution (GSD) generated by volcanic explosions is related to the efficiency of the magma fragmentation (e.g. Zimanowski et al., 2003; Kueppers et al., 2006; Cashman and Scheu, 2015; Liu et al., 2015) and provides crucial information about the mechanisms operating during an eruption. Several studies have focused on the size distribution of volcanic particles as a tool for understanding eruption dynamics and tephra dispersion. For example, Fisher (1964) determined wind directions during transportation of particles based on their medium diameters; Walker (1973) proposed a classification for explosive volcanic events based on the area of dispersal and the degree of fragmentation of the material; Koyaguchi and Ohno (2001) and Girault et al. (2014) used total grain size distribution (TGSD) of volcanic deposits to infer the dynamics of volcanic plumes.

Different types of statistical distributions have been used to characterize the GSD of volcanic deposits: Sheridan (1971) evaluated the mechanism of transport and deposition of volcanic particles and interpreted size-frequency distributions considering log-normal distributions; Murrow et al. (1980) and Spieler et al. (2003) used the Rosin-Rammler distribution to approximate the TGSD of natural tephra samples and experimentally generated pyroclasts; Nakamura et al. (2007) implemented the Weibull distribution to characterize GSDs of basalt; this distribution was later used by Gouhier and Donnadieu (2008) to explain the size distribution of volcanic ejecta. Sheridan et al. (1987) and Eychenne et al. (2012) identified polymodal grain-size distributions and interpreted them as due to sequential fragmentation during transport and deposition.

Several studies on fragmentation have shown that rock fragment size distributions can be approximated by a power-law distribution (Turcotte, 1986, 1989; Kaminski and Jaupart, 1998). This implies that scale invariant processes acted to generate the observed GSDs, pointing to the usefulness of fractal statistics as a tool to quantify the fragmentation mechanisms and obtain new insights on the physical processes that generated them. In particular, the scale-invariant behavior of GSDs due to magma fragmentation attracted the attention of volcanologists as a possible additional tool for assessing volcanic hazard. For example, Kueppers et al. (2006) and Perugini and Kueppers (2012) used fractal statistics to link GSDs to the energy of fragmentation of experimentally generated pyroclastic fragments. Kaminski and Jaupart (1998) and Jones et al. (2016) linked changes of fractal dimension of fragmentation to the occurrence of secondary fragmentation processes. Pepe et al. (2008) and Perugini et al. (2011) applied the fractal fragmentation theory to data from natural pyroclastic deposits in order to infer the evolution of eruption dynamics. The above experimental and field studies demonstrated that size distribution of fragmented volcanic materials can be approximated by a fractal distribution, suggesting that power-law models are well suited for describing volcanic fragmentation. Despite the documented suitability of fractal statistics in characterizing and quantifying volcanic GSD, several issues remain on how to use these methods to provide insights into the mechanisms acting in volcanic systems during the fragmentation process.

In this work, we use fractal statistics to investigate the fundamental mechanisms acting during magma fragmentation in the course of explosive volcanic eruptions by studying GSDs of natural pyroclastic-fall deposits. The method is applied to study lithic-rich proximal deposits from a representative stratigraphic section of the Cretaio Tephra eruption (Ischia Island, Italy). We collect GSDs for the Bulk Material (BM), Juvenile (JV), and Lithic (LC) fraction from each tephra layer and apply fractal fragmentation theory. Different scaling behaviours, characterised by different values of fractal dimension of fragmentation ($D$), are identified for different granulometric size ranges. Results are discussed as to what processes may have acted in the volcanic conduit in order to generate the different scaling regimes.

## 2. Stratigraphy and sampling

The Cretaio Tephra (1860y BP; Orsi et al., 1992) is a small-volume (< 0.02 km$^3$) fallout deposit generated by a vent located on the island of Ischia, belonging to the Phlegrean Fields volcanic district, Southern Italy (Fig. 1). The deposit was produced by a magmatic eruption and by minor initial and final phreato-magmatic phases. It is composed of juvenile pumice, ash and lithic clasts and it is distributed over the eastern sector of the island. We sampled material from a proximal section located at the locality of Cretaio, ca. 1 km from the proposed vent area (de Vita et al. 2010; Fig. 1). According to Orsi et al. (1992) and de Vita et al. (2010) the stratigraphic section can be divided into six Eruptive Units (EU): EUA, EUB, EUC, EUD, EUE and EUF. The lithological features observed in units EUC and EUD allowed us to further divide them into sub-units. Sub-units within EUC were divided based upon cycles of normally graded particles. Each cycle increases in thickness towards the top of the main unit (Fig. 2), indicative of a longer lasting pulse, or more material erupted. With regards to EUD, the division into 3 sub-units was achieved based on the distinct change in grain size and colour from one layer to the next. Stratigraphy and characteristics of eruptive units are represented in Fig. 2.

EUA corresponds to the basal bed, identified as an ash-surge deposit (Orsi et al., 1992 and de Vita et al., 2010), resting on a mature purple-brown paleosoil (Fig. 2). It has a grey-greenish colour, and contains lapilli-size juvenile clasts coarsening upwards and abundant lithics (ca. 55 wt.%). A coarse ash level, with particles of 0.5 to 1 mm in size, is observed in the central part of the unit. Finer grained material (< 63 μm) constitutes 5.8 wt.% of the deposit. EUB is a massive layer with brown-pinkish colour. Compared to EUA, it shows lower amounts of lithic clasts (ca. 39.2 wt.%), is better graded and shows a higher content of fine material (< 63 μm; ca. 14.2 wt.%) in which pumice clasts are embedded. The top of this unit is irregular, showing U-shaped erosional surfaces. EUC is a fallout deposit that we subdivided into 5 sub-units (EUC1 to EUC5) (Fig. 2). Each of these is composed of about the same percentage of white pumice fragments and grey/black lithic clasts (ranging from ca. 38.5 to 54.9 wt.%), and shows a normal gradation (Orsi et al., 1992). Ballistic clasts showing an ashy patina are present at the base of sub-unit EUC2. Fine material for the whole EU is in the range 0.48 to 2.83 wt.%. EUD was divided into three sub-units: EUD1 and EUD3 correspond to massive fall lapilli deposits, whereas EUD2 is a thin layer rich in ash particles. The wt.% of lithics and fine material (< 63 μm) for sub-units at EUD is quite variable. EUD2 shows the highest concentration of lithics and finer material, represented by ca. 60.1 wt.% and ca. 7.5 wt.%, respectively. Sub-units EUD1 and EUD3 show 18.3 wt.% and 34.7 wt.% of lithics, respectively; the content of fine material for both these sub-units is similar, i.e. ca. 2.0 wt.%. Lastly, EUF is fine ash-rich deposit (ca. 8.2 wt.%) with an abundant lithic content (ca. 57.0 wt.%). This member is overlaid by a developing soil. For more details on the lithological description of EU of the Cretaio eruption the reader is redirected to the work of Orsi et al. (1992).

In this work we focus on EUC and EUD, which correspond to the magmatic phase of the eruption. We do not consider EUF, as it is topped by a soil and therefore erosion processes and bioturbation might have perturbed the original grain size distribution. Five samples were collected from the EUC and three from the EUD, and analysed as reported below.

## 3. METHODOLOGY.

### 3.1. Sample preparation and grain-size analysis

Samples were placed in the oven at 60 °C for 48 hours to remove moisture. The bulk material (BM) was separated into two categories: the juvenile fraction (JV), defined as the pumiceous-vesiculated material, and the lithic fraction (LC), defined as the dense rock. This physical separation of the components was done through the winnowing procedure, as described by Gualda et al. (2004). To reduce the loss of the fine fraction during this process, the material was sieved at 125 μm (3 Φ) prior to physical separation. As a consequence, in this work we analyzed grain sizes larger than 125 μm. Resultant samples were analyzed under a binocular microscope with a magnification of 10x to check for the accuracy on the separation of lithics and juveniles. Density values for JV and LC were calculated by Dynamic Image Analysis (DIA) by means of Camsizer® P4, based on the known mass value and the total volume of the sample determined by image processing. The GSD of the eight stratigraphic layers were determined for the BM, and JV and LC fractions, separately.

*3.2 Dynamic Image Analysis (DIA)*

For each layer, a mass of ca. 250 g was used in the analyses (except for EUD2, which thickness was considerable lower than the rest of the sub-units). The GSD of each sample was measured using Dynamic Image Analysis (Goossens, 2007; Andronico et al., 2008, 2013). Dynamic Image Analysis (DIA) provides robust statistics of the measured quantities since particles are characterized in a dynamic process and in different orientations (Bagheri, 2015). In particular, we used the CAMSIZER® P4, a compact laboratory instrument for simultaneous measurement of particle size distribution by digital image processing (Retsch Technology GmbH Laboratories, Haan, Germany). The sample was fed-in from a vibratory feed channel that controls particle falling through the measurement field. The software controls the speed at which the sample reaches the measurement area. Then the time of vibration (ca. 3 min, on average) and the amount of the sample falling into the measurement field is set to avoid particle aggregation during the measurements. With the DIA technique, two-dimensional projections of particles are captured with two digital cameras, as they fall through the backlighted measurement volume. The two digital cameras cover the following size ranges: (1) a basic camera (CCD-B) registered the large particles (size range 625 μm ‒ 30 mm) whereas (2) a zoom camera (CCD-Z) recorded the smaller ones (size range 30 μm – 1250 μm). The overlapping of both cameras assured high accuracy on the results for a preferable size range between 625 μm and 1250 μm. A LED strobe light source (90 Hz) improved the brightness and the contrast of the images. The projected particle shadows were recorded at a rate of 60 frames per second. These were processed through the software CAMSIZER - Retsch Technology. We checked for possible secondary fragmentation generated by falling impact by repeating measurements on the same sample several times (typically three times). Results indicate that during measurements minor secondary fragmentation occurred generating an average increase of the ash fraction of ca. 0.8 wt.%.

From the image of each particle, the software measured its length and width. Then, assuming the geometric shape of an elongated rotational ellipsoid, it calculated the volume of each particle. Results were delivered as volume percentage (vol.%) on a discrete range of particle sizes with bins of approximately 0.2 Φ. We merged several consecutives fractions to obtain a consistent 0.5 Φ scale from -4.5 to 3 Φ, assuming that the volume percentage varies linearly in a 0.2 Φ interval of the results from DIA.

Results from DIA were converted to wt.%-based GSDs (Fig. 3), assuming constant density for the examined size range. The volume-to-mass conversion was obtained by using the mean value of density determined by the CAMSIZER®. The median values of density used for BM, JV and LC fractions are: 1.25 ± 0.10; 0.81 ± 0.02 and 1.86 ± 0.10 g/cm³, respectively for EUC. As for EUD, the median values of density used for BM, JV and LC fractions are: 1.13 ± 0.19; 0.80 ± 0.07 and 1.17 ± 0.06 g/cm³, respectively.

### 3.3. Fractal Fragmentation theory and size distribution of pyroclastic deposits

Fractal analysis has been applied to describe a wide variety of natural fragmented materials (Turcotte, 1986; Barnett, 2004; Pepe et al., 2008). Mandelbrot (1982) first showed that the fractal dimension for a given population of particles could be measured as:

$$N(r > R) = R^{-D} \qquad \text{(Eq. 1)}$$

where $D$ is the fractal dimension of fragmentation and $N(r > R)$ is the total number of particles with a linear dimension $r$ greater than a given comparative size $R$. Eq. 1 is a power-law size distribution from which the value of $D$ can be derived. $D$ is a measure of the relative abundance of fragments of different sizes or, in other terms, the degree of fragmentation of the population. Taking the logarithm of both sides of Eq. 1 yields a linear relationship between $N(r > R)$ and $R$, with $D$ representing the slope coefficient:

$$log[N(r > R)] = -D \, log(R) \tag{Eq. 2}$$

Tyler and Wheatcraft (1992) and Turcotte (1986, 1992) developed a more appropriate "mass-based" method where mass measurements can be directly used as:

$$\frac{M(r < R)}{M_0} = (R)^v \tag{Eq. 3}$$

where $M(r < R)$ is the total mass of fragments with a linear dimension $r$ less than a specified value $R$, $M_0$ is the total mass of particles, $R$ is the sieve size opening, and $v$ is a constant exponent. It is possible to derive a direct relationship between the power exponent $v$ and the number-based $D$ value given in Eq. 1. Taking the derivatives of Eq. 1 and 3 with respect to the size $R$, yields to the following relationships:

$$dN \propto R^{-D-1} dR \tag{Eq. 4}$$

$$dM \propto R^{v-1} dR \tag{Eq. 5}$$

The volume of a particle of size $r$ is proportional to its mass $m$, as $r^3 \propto m$. Thus, the incremental particle mass is related to the incremental particle numbers by:

$$dM \propto R^3 dN \tag{Eq. 6}$$

Substitution of Eq. 4 and 5 into Eq. 6 gives:

$$R^{v-1} \propto R^3 R^{-D-1} \tag{Eq. 7}$$

from which it follows that

$$D = 3 - v \tag{Eq. 8}$$

Therefore, $D$ can be calculated using the power law exponent $v$ from the mass-based approach. The later is found by estimating the slope of the linear fitting of log[M(r < R)] vs. log(r) (i.e. the cumulative frequency vs. size of the particle).

The analyses explained above were performed on a data set of 8 samples, including BM, JV and LC fractions. The $D$ values have been estimated using Eq. 8, by identifying the slope of the linear fitting of log[M(r < R)] vs. log(r) (i.e. the cumulative frequency vs. size of the particle).

## 4. RESULTS

*4.1 Grain size distributions of the fall deposits*

Approximate GSDs descriptions of the pyroclastic samples have been obtained using the parameters defined by Inman (1952): the median grain-size (Md$_\Phi$), the sorting ($\sigma_\Phi$) and the skewness (Sk) (Table 1). The results of the grain-size analysis are discussed with respect to the analyzed fractions (BM, JV and LC).

Bulk material (BM)

Values of $\sigma_\Phi$ for subunits on EUC do not vary significantly along the sequence (i.e. ca. 1.5 ± 0.16), indicating that deposits are moderately sorted (Fig. 3). Values of $\sigma_\Phi$ for EUD sub-units indicate a slightly better sorting ($\sigma_\Phi$ ca.1.2 ± 0.14, Table 1). EUD2 represents an exception, with a behavior slightly bimodal and a principal mode at around -2.5 Φ (5.6 mm), smaller than for EUD1 (ca. -3 Φ, 8.0 mm), and a smaller mode at around 3 Φ (125 μm).

Md$_\Phi$ values range between -3.5 and -2.4 Φ; these end-member values correspond to EUC2 and EUD2 respectively. The rest of the sub-units show relatively constant values of Md$_\Phi$, around approximately -3.0 Φ. Finally, Sk values are in the range 0.1 to 0.3 for most samples, with the exception of EUD3 that is more symmetrical, with a Sk value of 0.08 (Table 1).

Juvenile fraction (JV)

Moving to the top of EUC, GSDs exhibit a progressive change from relatively good to moderately sorted (Fig. 3). At the same time, sorting of unit EUD shows a slight decrease towards the top of the unit. σ$_\Phi$ values are between 0.8 and 1.3 for EUC, and from 0.9 to 1.1 for EUD. As observed for BM, EUD2 exhibits a bimodal character with two maxima at -2.5 Φ (5.6 mm) and 3 Φ (125 μm).

Md$_\Phi$ values range between -3.2 and -3.8 Φ, except for EUD2 showing the lowest value of the entire sequence (-2.71 Φ). Sk values are in the range 0.2 and 0.4, with the exception of sub-members EUC2, EUD1 having Sk values close to zero (ca. -0.05 to 0.02), and consequently more symmetrical distributions (Table 1).

Lithic fraction (LC)

The sorting of GSDs for LC samples, in general, does not correlate with stratigraphic height for EUC and EUD and values of σ$_\Phi$ are nearly constant (1.1 and 1.6 Φ) indicating moderately sorted distributions. These distributions exhibit lower modes relative to GSDs for JV from both EU. Md$_\Phi$ values range between -1.8 to -2.6 Φ; Sk values range from 0.02 to 0.25. As for EUC, sub-members EUC1, EUC3, EUC4 and EUC5 have similar symmetrical distributions, whereas EUC2 shows a more skewed distribution (Sk ca. 0.25, Fig. 3). In EUD, the above three parameters (σ$_\Phi$, Md$_\Phi$ and Sk) are quite similar for the three sub-units.

*4.2 Fractal statistics*

The GSDs of the samples from the fall deposits were evaluated for their fractal character (Fig. 4). The fractal analysis can: i) determine the fractal dimension of fragmentation (*D*) of the deposits, and ii) identify possible correlations between the different *D* values for BM, JV and LC, in order to unveil potential relationships that might aid in understanding eruptive dynamics.

The plots presented in Figure 4 highlight that a single power law relationship can be used to represent the GSDs of bulk material, but fails to explain GSDs for JV and LC. In particular, JV and LC grain size distributions appear to be characterized by two power laws with different *D* values for all the sub-members of the sequence. Previous studies reported the presence of more than one scaling regime on natural samples and experiments (Hatton et al., 1994; Main et al., 1999; Grady, 2010; Costa et al., 2016). Accordingly, we fitted the BM grain size distributions using one power-law relationship, whereas two power-law relationships were used to fit the GSDs of JV and LC (Fig. 4).

The plots of Figure 4 strongly support two scaling regimes for JV and LC; however identification of the threshold values for the two scaling regimes (i.e. the two slopes in the log-log plots) is not trivial. Some methods have been developed to determine the changes in the scaling regime for scale-dependent distributions (Main et al., 1999; Grady, 2008; 2010); for example Hatton et al. (1994) propose that the change between two distributions can be determined by simple observation; then, the degree of confidence in defining the different distributions is provided by the value of standard deviation of the relative slopes. Given that there is not a general consensus about which method has to be used to identify contiguous distributions in a dataset, in this work we used the following procedure (Fig. 5): once a kink in the slope was observed in the plot, the data point corresponding to it was selected and associated to one possible linear trend (Fig. 5A). Subsequently, additional data points were added to this linear trend (to the left or the right of the kink) and a linear fitting performed in the log-log space (Fig. 5B). Correlation coefficients ($r^2$) and the standard deviation ($S$) resulting from the linear fitting were estimated. The kink was considered valid when $r^2$ values were greater than 0.9 and the standard deviation ($S$) was equal or lower than $\pm 1.0 \times 10^{-1}$ (Fig. 5C and Fig. 5D). As for BM no clear kink in the GSD was observed, consequently, the distribution was fitted considering a single linear relationship.

Bulk Material (BM)

Power-law fitting of BM datasets provides values of $D$ in the range 1.84 and 2.20 (Table 2). Due to the erratic distribution of the data points for sample EUD3, the fitting couldn't be satisfactorily performed on BM, JV and LC data. Noteworthy is the fact that this member is overlaid by the base surge deposit EUE that might have perturbed the original GSD. Accordingly, this sample will not be included in the discussion below.

Juvenile fraction (JV)

Two scaling regimes characterize the grain size distribution for all the JV samples, with the kink in the range 1.0 – 2.83 mm, with most of the samples having the kink at 2.0 mm. These sub-fractions were defined as "$JV_A$" (left side of the kink, i.e. the finer fraction) and "$JV_B$" (right side of the kink, i.e. the coarser fraction) (Fig. 4).

Results from the fitting indicate that the slope ($v$) for the "$JV_A$" fraction is lower than the slope ($v$) for the "$JV_B$" fraction for all eruptive units. According to Eq. 8, this condition reflects a higher value of $D$ for the finer fraction relative to the coarser one [$D(JV_A) > D(JV_B)$] (Table 2).

Lithic fraction (LC)

All the members show a very similar behavior in their fractal character, with one scaling regime extending from the finest sizes to 4.0 mm. The only exception is sample EUD2 for which the finest fraction extends up to 2.0 mm. The kink highlights the presence of a second linear distribution extending towards the coarser sizes. We name the GSD fitted by the first linear relationship (i.e. the finer fraction) as "$LC_A$", and the second linear relationship fitting the coarser fraction as "$LC_B$" (Fig. 4).

Results from the linear fitting procedure indicate that, contrary to the JV fraction, value of the slope ($v$) for "$LC_A$" is higher than the one for "$LC_B$" for all the eruptive units (Table 2). According to Eq. 8, this condition reflects a lower value of $D$ for the finer fraction relative to the coarser one [$D(LC_A) < D(LC_B)$] (Table 2).

Influence of end-member data points on the estimated values of D

Visual inspection of the graphs in Fig. 4 shows the presence, in some cases, of end-member data points, placed on the left and/or right side of the plots, that slightly deviate from the liner trends. In order to test whether these data points might influence the slope values and, consequently, the estimated values of fractal dimension of fragmentation, we estimated $D$ values by removing these points. Results indicate that the values of fractal dimension do not vary significantly and they remain within the errors of $D$ reported in Table 2. Further information can be found in the Supplementary Material.

5. DISCUSSION

Figure 6 shows the variation of fractal dimension of fragmentation ($D$) along the pyroclastic sequence for bulk material (BM), and for finer and coarser portions from the juvenile ($JV_A$ and $JV_B$) and lithic ($LC_A$ and $LC_B$) fractions.

As for the BM, the fact that grain size distributions can be quantified using a single power-law argues in favor of the hypothesis that a single fragmentation mechanism might explain the observed GSDs. The variation of $D$ values along the pyroclastic sequence display nearly constant values. This might be considered as an indication that in all eruptive units the degree of fragmentation was essentially similar as the eruption developed in time.

Contrary to BM, lithic and juvenile fractions show two scaling regimes, thus reflecting the possible action of different fragmentation mechanisms/events that acted to generate the observed GSDs. In particular, the finer and coarser particle size ranges for LC and JV ($LC_A$-$LC_B$, $JV_A$-$JV_B$, $LC_A$-$LC_B$) display opposite behavior as indicated by the fact that $D(LC_A) < D(LC_B)$ and $D(JV_A) > D(JV_B)$. In addition, the variation of $D$ along the sequence for the juvenile ($JV_A$ and $JV_B$) and the lithic ($LC_A$ and $LC_B$) fractions are characterized by statistical variations of this value from the bottom to the top of the sequence (Fig. 6).

According to the above discussion, it appears therefore clear that, depending on the type of fraction that is analyzed (BM, JV or LC), different and controversial information can be derived from the GSDs. In the following discussion we attempt to link the observed complexity of grain size distributions to the possible processes that, during eruptive activity, might have operated to generate the different scaling regimes. In order to do this, the basic mechanisms developing fractal size distributions need to be taken into account.

The conceptual model used to derive Eq. (3) and (5) is based on the self-similar fragmentation of a mass into progressively smaller particles (Matsushita, 1985; Turcotte, 1986). In particular, fragmentation starts from a cubic shape of size $h$ and fragments into eight smaller cubes of size $h/2$. These smaller cubes are further fragmented following an iterative procedure to give cubes with size $h/4$, and so forth. In this fragmentation model, the cube has a certain probability, $p$, to be fragmented, which is assumed to be constant for all orders of fragmentation. The cube is maximally fragmented into eight smaller cubes if $p = 1.0$ and into one smaller cube if $p = 1/8$. Turcotte (1986) established the following relationship between the fragmentation probability, $p$, and fractal dimension of fragmentation, $D$:

$$D = \frac{\log(8p)}{\log(2)} \tag{Eq. 9}$$

with the possible range of fractal dimension being $0 < D < 3$. A plot of Eq. (9), along with $p$ values calculated using values of $D$ measured on the studied samples, is given in Fig. 7, where results for BM and LC are compared with results from BM.

The fractal analysis performed on the lithic fraction shows that the probability of fragmentation ($p$) for the coarser particles is always larger than $p$ for the small particles $p(LC_A) < p(LC_B)$. The opposite behavior is observed for the juveniles, where $p(JV_A) > p(JV_B)$. Note that $p$ for the bulk material has intermediate values between those of LC and JV.

Considering the above results and discussion, the question arises as to what processes might have been responsible for this variability of $D$ and $p$ values in the different size fractions for the studied pyroclastic sequence. A first issue concerns the presence of a single scaling regime (i.e. one value of fractal dimension of fragmentation, $D$) observed for the bulk material (BM), which reflects a single value of probability of fragmentation ($p$). This behavior for BM has to be considered as the combination of GSDs belonging to both the juvenile and lithic components. This being so, the BM is not likely to provide direct information upon the mechanisms acting in the volcanic system and triggering magma fragmentation. In fact, this information can be hindered by the mixture of JV and LC components, possibly denying the possibility to derive information about eruption mechanisms starting from the analysis of the BM. This aspect is particularly relevant for the studied pyroclastic sequence because it is a proximal deposit characterized by high percentages of lithics (18.3 – 60.1 wt.%). When the amount of lithics is low, for example in lithics-poor eruption or distal deposits this problem is expected to be minimized. Noteworthy is the fact that, basing the interpretation of eruption dynamics solely upon the variation of $D$ along the pyroclastic sequence for the BM, this value appears constant (Fig. 6A). This could be misinterpreted as the eruption being constant in character through time. Furthermore, because the fractal dimension of fragmentation is proportional to the energy available for fragmentation (e.g. Kueppers et al., 2006; Perugini and Kueppers, 2012) one might be erroneously tempted to infer that the eruption energy available for fragmentation did not change during the course of the eruption. However, in BM resulting from mixing of JV and LC particles, with a large percentage of LC, this one cannot be taken as fully representative of the mechanisms that acted during magma fragmentation.

As for the variation of fractal dimension of fragmentation of JV and LC, they might potentially carry different information about the fragmentation processes that generated them. In fact, the JV populations arise from fragmentation of the magma, due to a decompression event, and their sizes and shapes should be always sensitive indicators of eruptive processes (Carey and Houghton, 2010). Lithics, on the contrary, are rock particles derived from the walls of the volcanic conduit and/or the vent that can be incorporated into the erupting mixture by conduit wall disruption/implosion around and above the fragmentation depth, shallow vent spalling (Hanson et al., 2016), and the passive entrainment of already loose wall rock by the gas and pyroclast mixture (Houghton and Nairn, 1991). In the case of pyroclastic deposits in which large amount of lithics are present, the possible interaction between juvenile fragmented material and the lithic component has to be taken into account. In particular, it is likely that the presence of lithics, if entrained deep in the conduit, might have played a role in the re-fragmentation of the JV component during the eruption.

In the following we attempt to explain the different populations and scaling regimes observed for JV and LC fractions according to two different fragmentation mechanisms.

*5.1 Fragmentation of the lithic fraction*

The results for the lithic fraction display a probability of fragmentation ($p$), which decreases as the size of particles forming the population decreases [$p(LC_A) < p(LC_B)$] (Fig.7A). The two linear relationships observed on the plots of Fig. 4 suggest the presence of two distinct fragmentation mechanisms that acted at different length scales. In particular, the fragmentation mechanism that operated on larger particles was more efficient than the one that generated the smaller particles [$D(LC_A) < D(LC_B)$]. This behavior can be explained, as a first approximation, considering the results that Carpinteri and Pugno (2002) obtained for fragmentation of solid materials. In this model, these authors consider the effect of energy dissipation during comminution processes leading to fractal GSDs. In particular, they observe different slopes for the same GSD when the fragmentation mechanism that operated on larger particles was more efficient relative to the one that generated the smaller particles, as for the behavior of the lithic fraction in this study. The different scaling regimes (i.e. different $D$ values) can be explained by the action of two main different processes operating at different length scales, associated to the same fragmentation event, corresponding to: (1) "cutting", acting at larger scales and leading to the formation of the larger particles, and (2) "milling", acting at smaller scales and leading to the formation of the smaller particles. According to Carpinteri and Pugno (2002), the different efficiency of fragmentation at the different length scales is related to the way energy dissipates. During the generation of the larger particles ("cutting") energy dissipation occurs substantially in the volume ($D$ values tending to 3), whereas the energy dissipation during the "milling" process mainly occurs on the surface area ($D$ values tending to 2). The different slopes in the GSDs for lithics in the studied pyroclastic sequence could be interpreted as the "cutting" process being the detachment of wall rocks (e.g. conduit wall disruption above the fragmentation depth, vent spalling, etc. Macedonio et al., 1994; Carey et al., 2007; del Gaudio and Ventura, 2008; Campbell et al., 2013; Eychenne et al., 2013) and the "milling" process being the processes occurring by continuous collision, grinding and friction of lithics against lithics or lithics against the conduit walls (Kaminski and Jaupart, 1998; Dufek et al., 2012; Campbell et al., 2013; Mueller et al., 2015; Jones et al., 2016; Bernard and Le Pennec, 2016; Jones et al., 2017). According to this model and results presented above, the kink in the fractal distributions (Fig. 5) indicates that 4 mm was the threshold particle diameter for our system, above which fragments were mainly generated by the "cutting" process. Under this particle size threshold particles mainly underwent "milling" processes.

*5.2 Fragmentation of the juvenile (JV) fraction*

As discussed above, the probability of fragmentation ($p$) increases as the size of particles forming the population decreases [$p(JV_A) > p(JV_B)$] (Fig. 7B). This scenario cannot be explained by existing fragmentation models leading to fractal size distributions. In particular, the presence of a more efficient mechanism of fragmentation acting on smaller scales (relative to larger scales) disagrees with fragmentation models where larger particles tend to fragment more easily than smaller ones (e.g. Perfect, 1997). Further, experimental fragmentation of natural pyroclastic rocks highlighted that the GSDs are characterized by fractal distributions with a single value of fractal dimension of fragmentation ($D$), i.e. a single scaling regime (Kueppers et al., 2006; Perugini and Kueppers, 2012) corresponding to a single event of fragmentation. This is in contrast to what is shown by the natural samples from the studied pyroclastic sequence, where multiple scaling regimes are observed (Fig. 4).

From the above discussion it is clear that different events of fragmentation, characterized by different efficiency, are needed to explain the scaling behaviour of the JV; these events must have taken place syn-eruptively, prior to juvenile deposition. Results from Dufek et al. (2012) could explain the behaviour observed on the JV fraction; their data indicated that sequential fragmentation of juveniles could occur in volcanic conduits, as a result of disruptive processes (i.e. collisions), leading to a significant variation of the initial population. Accordingly, the two scaling regimes observed in the plots of Fig. 4c (juvenile fraction) might represent two different fragmentation events: i) the initial fragmentation of the magma, evidences of which are possibly still preserved in the coarser fraction, and ii) the sequential fragmentation of the original GSD due to collisional events in the conduit. Bernard and Le Pennec (2016) showed that the efficiency of sequential fragmentation processes could be enhanced by the presence of large quantities of lithics. This is also confirmed by the experimental work of Kaminski and Jaupart (1998) indicating that an enrichment of lithics (analogous to the steel ball used in the experiments) in a magmatic eruption, could induce a more efficient "secondary" fragmentation, due to collisional events, leading to an increase of particles in the fine fraction. As discussed above, studied samples contain lithics contents in the range of 18 – 60 wt.%, possibly corroborating the idea that sequential fragmentation of juvenile particles might have been a highly probable process.

The kink in the juvenile fractal distributions at 2 mm (Fig. 4c) indicates that this particle diameter can be considered as the threshold value below which fragments were produced mainly by re-fragmentation due to collisional events. It follows that particle larger than 2 mm might still represent the GSD generated by the initial fragmentation of the magma. Therefore, the trend of $D$ values for the juvenile coarser fraction [$D(JV_B)$] along the eruption stratigraphy can be utilized to infer the possible time evolution of the eruption that generated the studied pyroclastic deposits. Recalling that the value of $D$ is a measure of the energy available for fragmentation (e.g. Kueppers et al., 2006), it can be said that the evolution of studied eruption, in the relative time window represented by the pyroclastic sequence, was characterized by a high-energy pulse represented by deposit EUC1, followed by a less energetic explosion, recorded in EUC2. Afterwards, the eruption energy reached a relative climax at EUC4 and then decreased to EUD1. The last studied deposit (EUD2) highlights a potential further increase of eruptive energy.

## 6. SUMMARY AND CONCLUSIONS

In this work we used fractal statistics of GSDs of natural pyroclast deposits to decipher the mechanisms acting during explosive volcanic eruptions. The method was applied to lithic-rich proximal deposits from a representative stratigraphic section of the Cretaio Tephra eruption (Ischia Island, Italy). The GSDs for the bulk material, juvenile, and lithic fraction from each pyroclastic layer yield different scaling behaviours, having different fractal dimensions (*D*) and different granulometric size ranges. In particular, we observed one scaling regime for the bulk material and two scaling regimes, with contrasting power-law exponents, for the juvenile and lithic fractions.

According to our interpretation, the following conclusions can be drawn:

a) in lithic-rich pyroclastic deposits the bulk material extracted from pyroclastic deposits cannot be considered as a proxy to infer the mechanisms acting during the development of an eruption and, therefore, it is not representative of eruption dynamics. In fact, the bulk material is a mixture of fragments belonging to the lithic and the juvenile fraction, both of which underwent different events of fragmentation, governed by different mechanisms;

b) the lithic fraction likely experienced a solid state fragmentation event, in which the coarser and finer fractions were subjected to different mechanisms that generated different particle size distributions at different length scales. This remained fossilized in the lithic GSDs as two different scaling regimes in which the efficiency of fragmentation was larger for the coarser fraction relative to the finer one;

c) the juvenile GSDs, after the initial explosion of the magma at the fragmentation level, underwent sequential events of fragmentation in the conduit, a process possibly enhanced by the presence of lithic fragments constituting the eruption mixture. Collisional events generated increasing amount of finer particles, modifying the original GSDs. The result, as observed in the studied outcrop, is the development of two scaling regimes in which the finer fraction was characterized by a higher efficiency of fragmentation relative to the coarser one;

d) possible indications about the original GSDs of the juvenile fraction might still reside in their coarser fractions. As such, this fraction might be used to reconstruct eruption dynamics and its time development during the course of the eruption.

The results and conclusions arising from this work need to be thoroughly tested through the study of additional natural pyroclastic deposits to check for the presence of different scaling regimes in the different components (lithics and juveniles) that might indicate the occurrence of the fragmentation events and dynamics suggested above. Along with the study of natural outcrops, new experiments need to be designed (for example using the fragmentation bomb; e.g. Kueppers et al., 2006) in order to understand how the presence of different amounts of solid material (lithics) can trigger sequential fragmentation of the juvenile material and the impact of these processes upon modification of the original grain size distribution. This might help in identifying whether threshold values of the amount of lithics exist, below which the juvenile grain size distribution can still preserve information about the initial grain size distribution.

Avoiding doing so, the use of natural pyroclastic deposits to infer eruption mechanisms and dynamics, especially in the proximal facies where the amount of the lithic fraction tends to be abundant, remains elusive.

**Acknowledgements**

This research was funded by the European Union's Seventh Programme FP7 Framework "FP7-PEOPLE-2013-ITN", under Grant agreement No. 607905 – VERTIGO and by the European Research Council for the Consolidator Grant ERC-2013- CoG Proposal No. 612776 CHRONOS to Diego Perugini. We thank the editor J.K.R., and two anonymous reviewers for their constructive comments that improved the manuscript, as well to Ph.D. student R. Astbury for proofreading of the manuscript.

**Additional information**

Supplementary data to this article have been provided. This material is intended to show the influence of largest- and smallest-scale data points in the estimation of $D$ values. Besides the GSD data for the complete section is presented, for each of the fractions analysed (i.e. BM, LC and JV)

**Figure Captions**

Figure 1. Location map (modified after de Vita et al. 2010) for the studied samples. A) Outline map of Ischia in the same volcanic province as the Phlegrean Fields; B) Ischia Island, showing location of measured stratigraphic section at the locality of Cretaio (blue star), vent

area proposed by Orsi et al. (1992) and dispersion area of the Cretaio Tephra deposits from de Vita et al. (2010).

Figure 2. Stratigraphic section sampled from the Cretaio Tephra eruption, from outcrop at the locality of Cretaio. The stratigraphic section was divided into six Eruptive Units (EU): EUA, EUB EUC, EUD, EUE and EUF, according to de Vita et al. 2010. EUC and EUD were further divided into sub-members (see text for details). Thickness and main features of the sequence sampled are indicated.

Figure 3. Representative grain-size distributions from the studied pyroclastic sequence for representative sub-units: EUC1, EUC2, EUC5 and EUD1. A) bulk material (BM); B) lithic (LC) and C) juvenile (JV) fractions.

Figure 4. Log-log representation of the cumulative frequency vs. size $\log[M(r < R)]$ vs. $\log(r)$ for same representative samples shown in Fig. 3. Linear fitting of GSDs are displayed for A) bulk material (BM); B) lithic (LC) and C) juvenile (JV) fractions. In the plots are reported the values of correlation coefficient ($r^2$), slope ($v$), and the corresponding value of fractal dimension of fragmentation ($D$). Plots for all analysed samples are reported in the Supplementary Material.

Figure 5. Illustration of the method used to identify the different scaling regimes for the GSDs of the studied samples (shown for the juvenile fraction of sample EUC1). A) Identification of the potential kink in the GSD; B) four consecutive data were fitted to a regression line, to the left (finer material, $JV_A$) and to the right (coarser material, $JV_B$) of the potential kink; $r^2$ and slope ($v$) values were calculated; C) additional data points were added to these linear trends (to the left and the right of the kink) and a linear fitting performed in the log-log space; D) $r^2$ values and the standard deviation ($S$) resulting from the linear fitting were estimated. The kink was considered valid when $r^2$ values were greater than 0.9 and the standard deviation ($S$) of slopes ($v$) was equal or lower than $\pm 1.0 \times 10^{-1}$

Figure 6. Graphs plotting the variation on fractal dimension of fragmentation ($D$) along the pyroclastic sequence, for: A) bulk material (BM); B) BM and lithic (LC) fraction and C) BM and juvenile (JV) fractions. The variation of D for BM is also reported in panels B and C for comparison.

Figure 7. Variation of probability of fragmentation ($p$) as a function of fractal dimension of fragmentation ($D$). Panels A and B show the comparison between BM and LC, and BM and JV, respectively.

**Table Captions**

Table 1. Calculated Inman's parameters (Inman, 1952) for the GSDs of the studied samples from the Cretaio pyroclastic sequence. Inman parameters were calculated for the BM, JV and LC fractions.

Table 2. Summary of the fractal dimensions of fragmentation ($D$) for BM, LC and JV fractions, for all the members on the pyroclastic sequence, excepting EUD3, which fitting couldn't be satisfactorily performed. The two values of $D$ for lithic and juvenile fractions represent to the two power-laws fitted for the entire distribution. Correspondent $r^2$ values for the fitting of the regression lines are displayed.

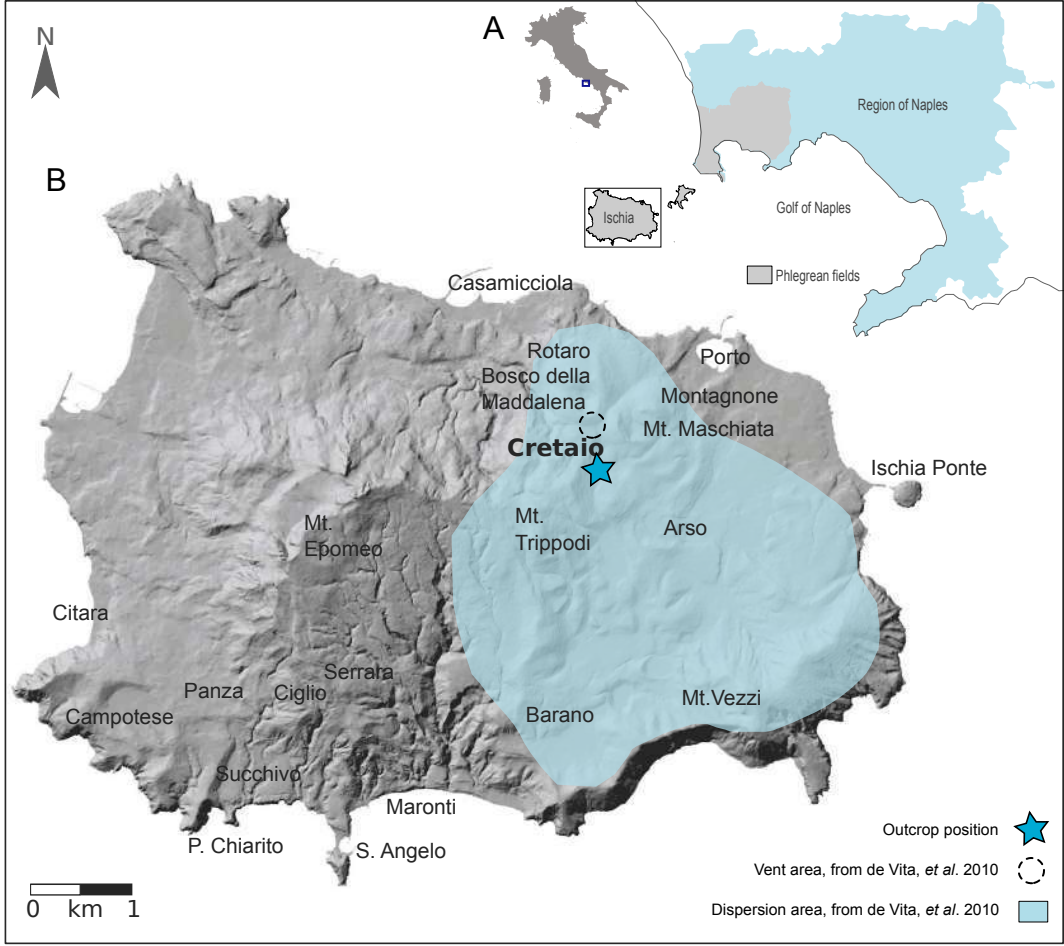

**Figure 01**

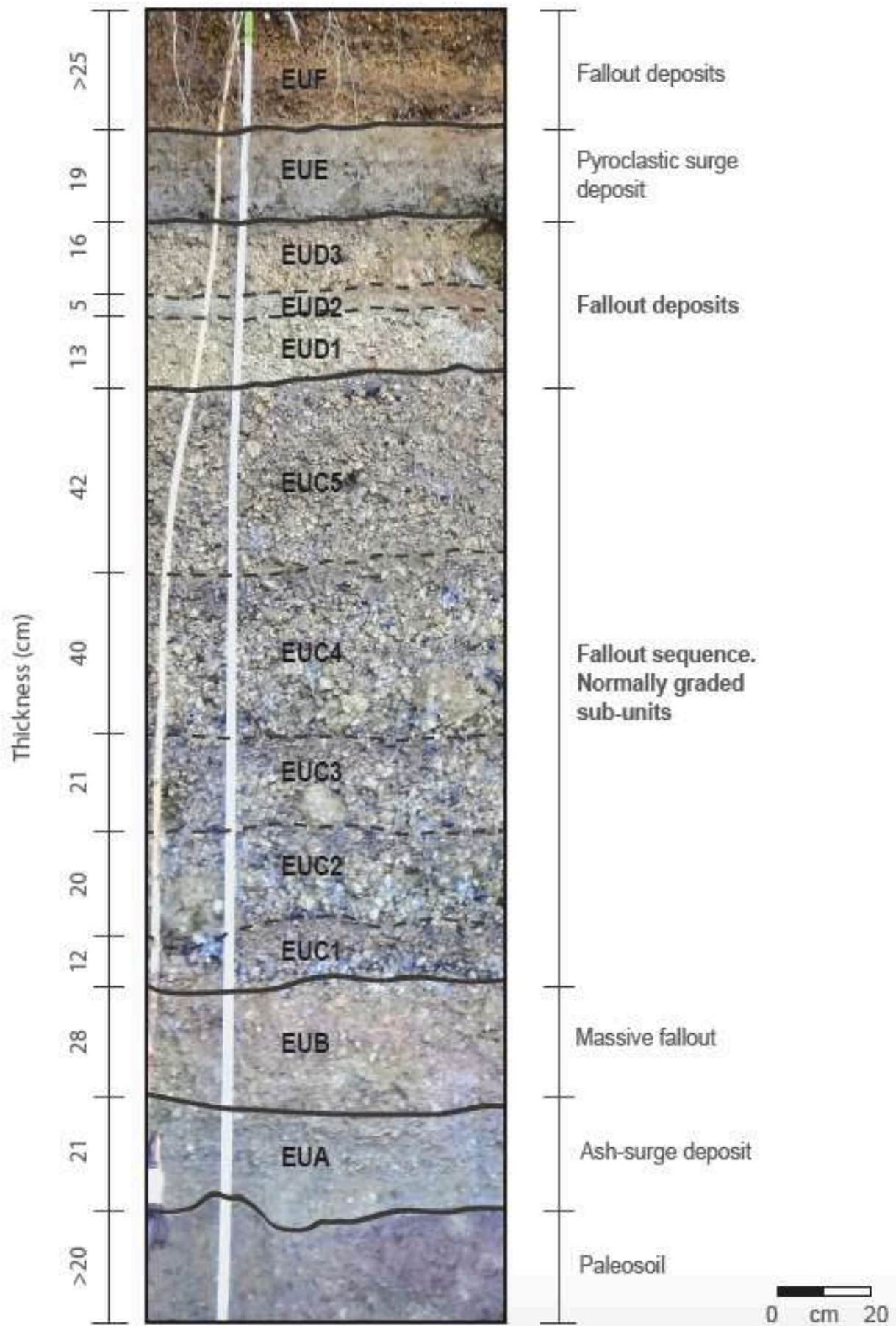

**Figure 02**

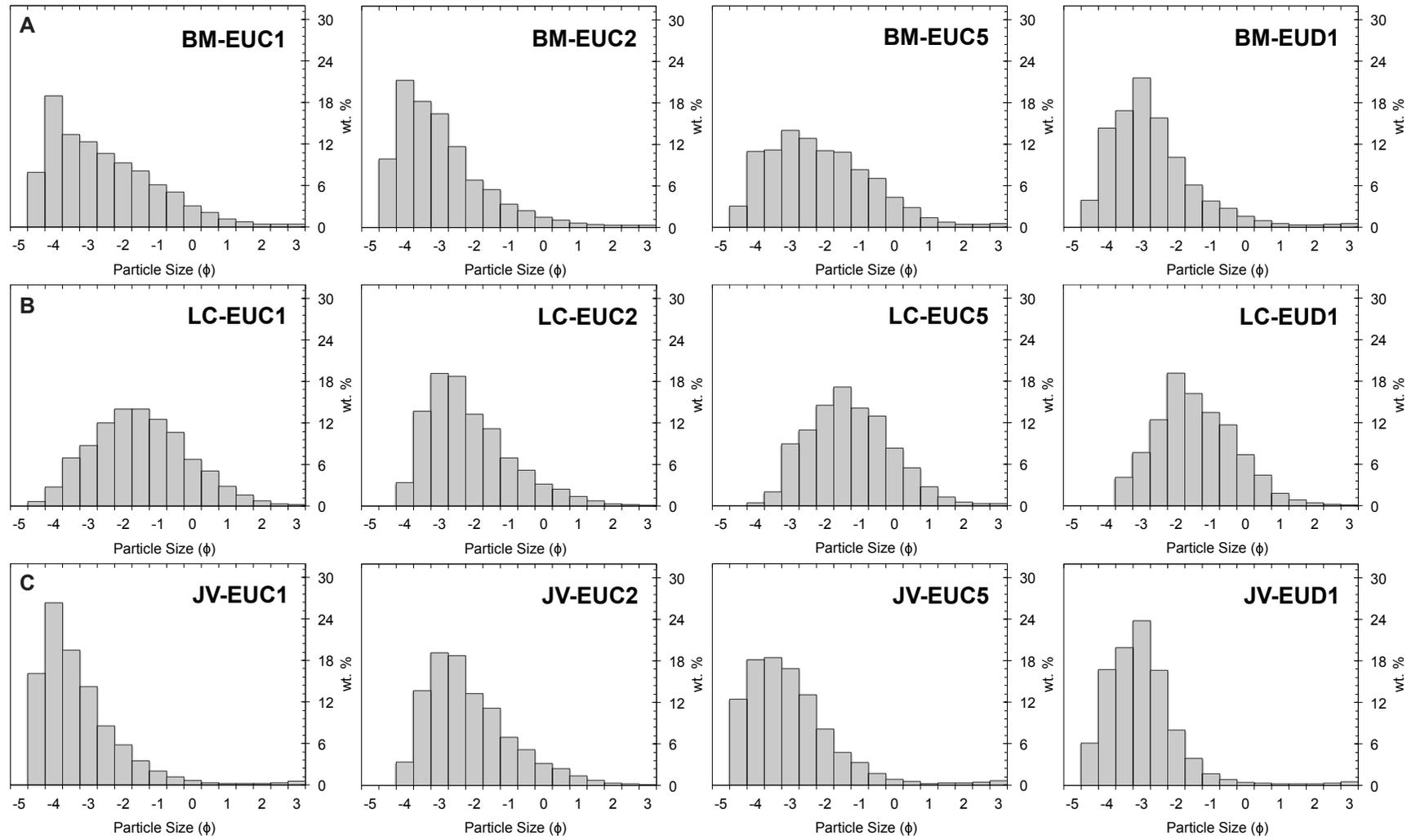

Figure 03

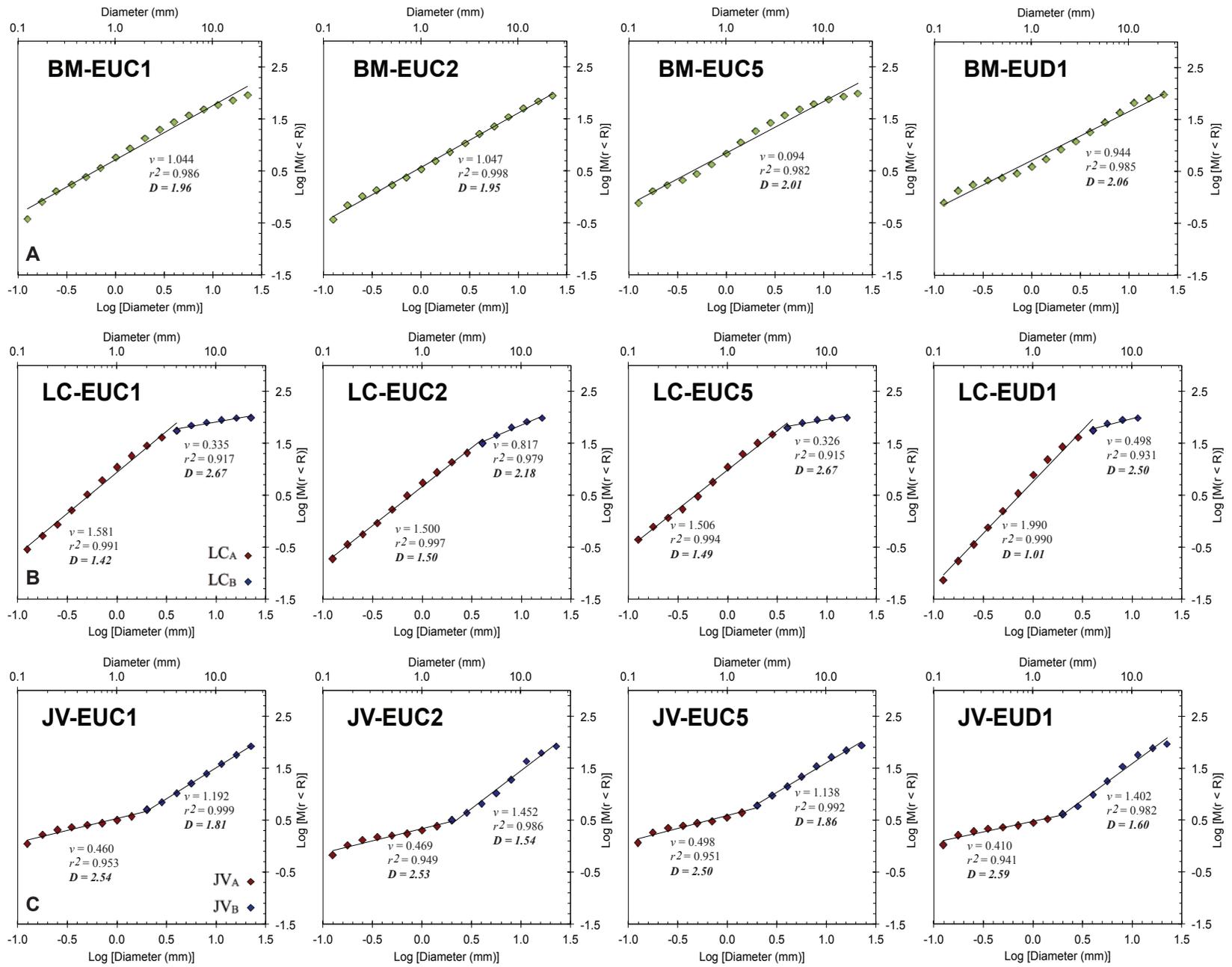

**Figure 04**

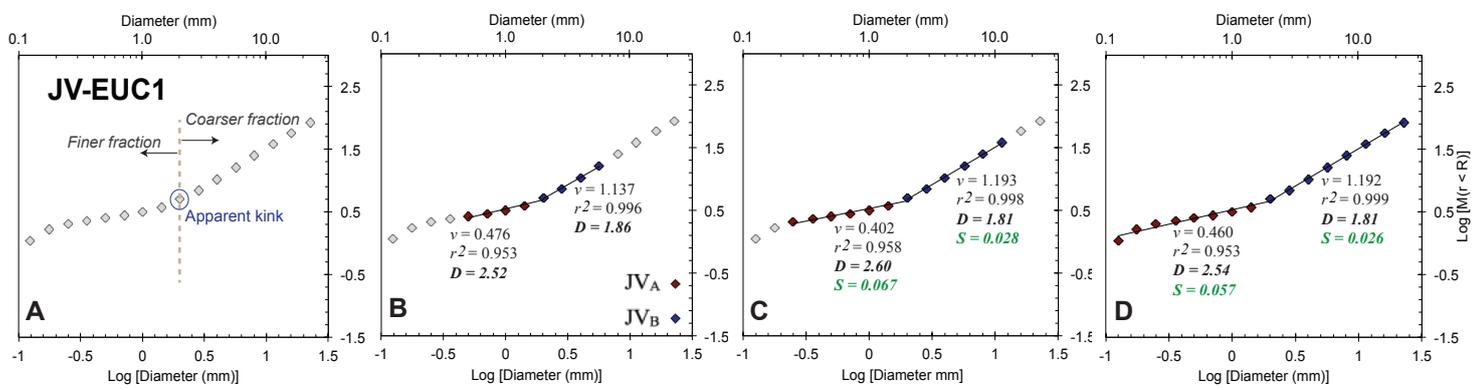

**Figure 05**

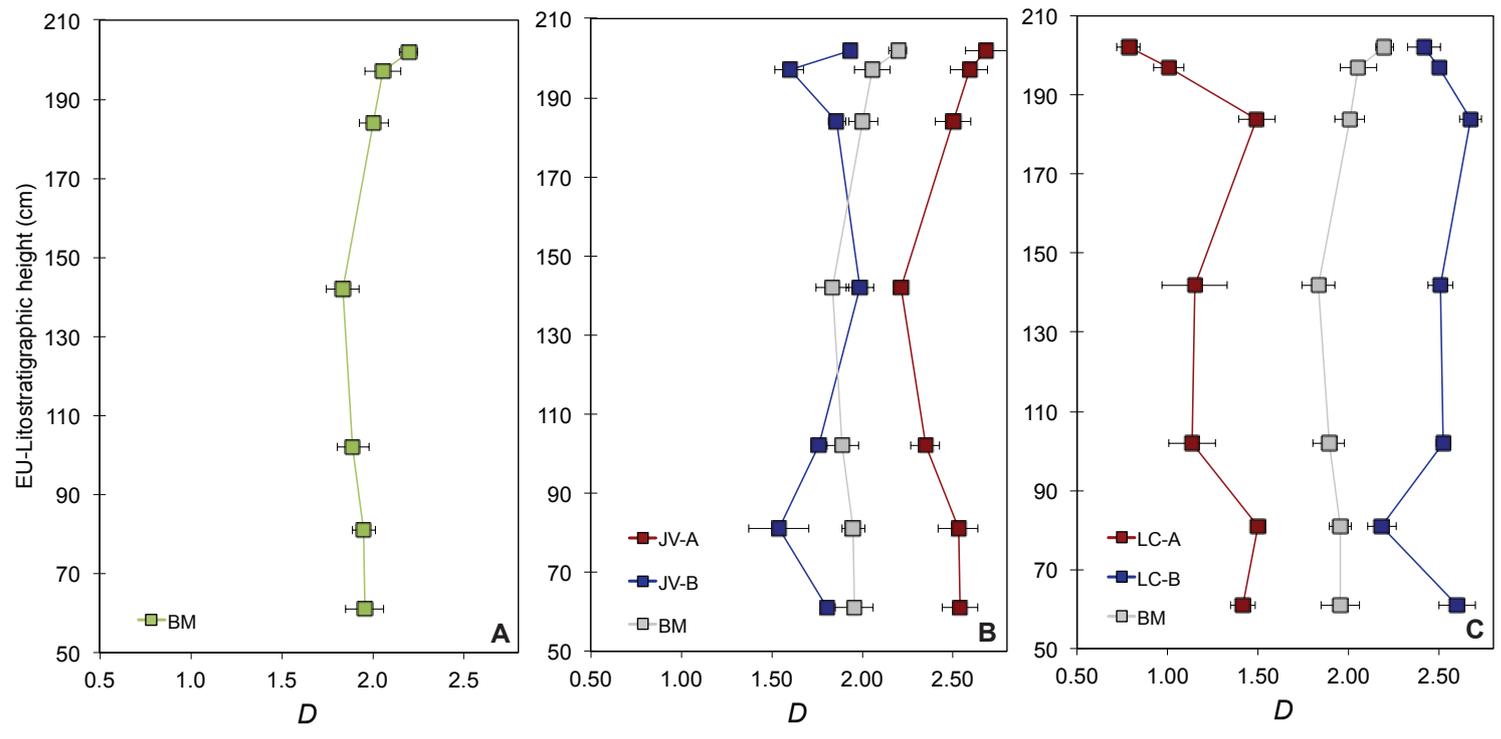

Figure 06

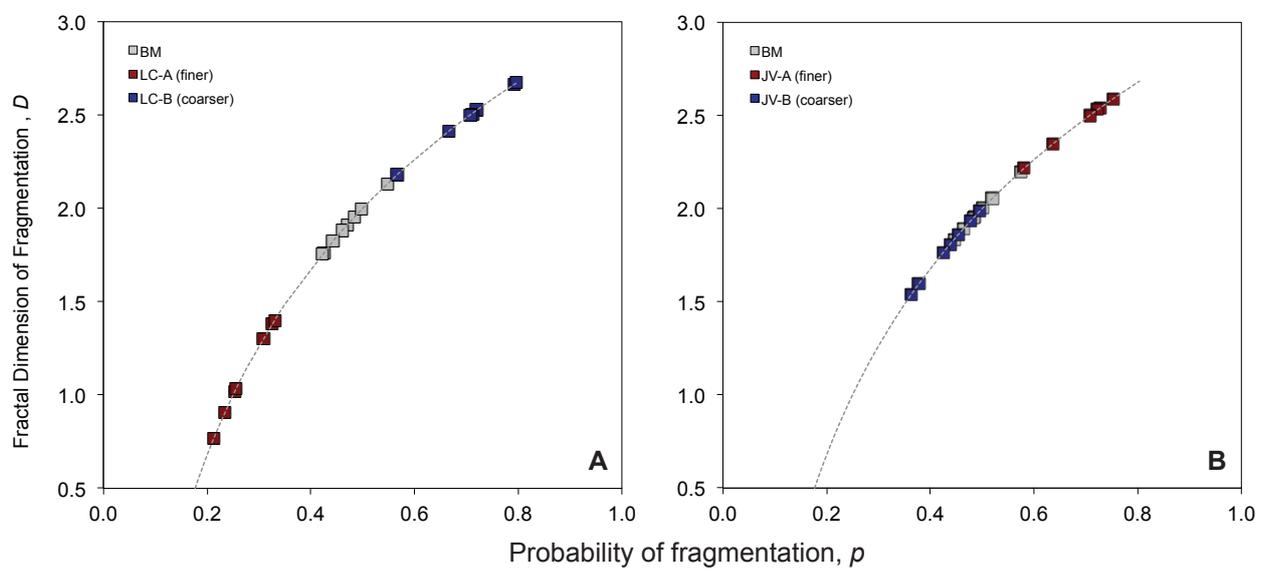

**Figure 07**

| Sample | Bulk material | | | JV-fraction | | | LC-fraction | | |
|---|---|---|---|---|---|---|---|---|---|
| | $Md_\Phi$ | $\sigma_\Phi$ | SkG | $Md_\Phi$ | $\sigma_\Phi$ | SkG | $Md_\Phi$ | $\sigma_\Phi$ | SkG |
| **EUD3** | -2.93 | 1.19 | 0.08 | -3.21 | 1.11 | -0.14 | -2.00 | 1.17 | 0.02 |
| **EUD2** | -2.42 | 1.36 | 0.32 | -2.71 | 0.95 | 0.32 | -1.88 | 1.39 | 0.13 |
| **EUD1** | -3.16 | 1.08 | 0.16 | -3.35 | 0.87 | 0.02 | -1.81 | 1.13 | 0.12 |
| **EUC5** | -2.58 | 1.51 | 0.12 | -3.47 | 1.10 | 0.15 | -1.62 | 1.22 | 0.05 |
| **EUC4** | -2.94 | 1.55 | 0.16 | -3.35 | 1.32 | 0.22 | -2.05 | 1.28 | 0.06 |
| **EUC3** | -3.17 | 1.56 | 0.32 | -3.75 | 0.96 | 0.35 | -2.14 | 1.56 | 0.04 |
| **EUC2** | -3.48 | 1.18 | 0.25 | -3.67 | 0.79 | -0.05 | -2.65 | 1.18 | 0.25 |
| **EUC1** | -3.11 | 1.54 | 0.23 | -3.82 | 0.98 | 0.30 | -1.83 | 1.38 | 0.04 |

**Table 1**

|  | Bulk material | | | Lithic fraction | | | | | | Juvenile fraction | | | | | |
|  | Whole size range | | | LC$_A$ (Finer) | | | LC$_B$ (Coarser) | | | JV$_A$ (Finer) | | | JV$_B$ (Coarser) | | |
| Member | D | $r^2$ | S | D | $r^2$ | S | D | $r^2$ | S | D | $r^2$ | S | D | $r^2$ | S |
|---|---|---|---|---|---|---|---|---|---|---|---|---|---|---|---|
| EUD2 | 2.20 | 0.995 | 0.049 | 1.04 | 0.987 | 0.095 | 2.48 | 0.949 | 0.083 | 2.68 | 0.919 | 0.114 | 1.94 | 0.994 | 0.028 |
| EUD1 | 2.06 | 0.985 | 0.099 | 1.01 | 0.990 | 0.083 | 2.50 | 0.931 | -- | 2.59 | 0.941 | 0.103 | 1.60 | 0.982 | 0.078 |
| EUC5 | 2.01 | 0.982 | 0.081 | 1.49 | 0.994 | 0.100 | 2.67 | 0.915 | 0.060 | 2.50 | 0.951 | 0.098 | 1.86 | 0.992 | 0.047 |
| EUC4 | 1.84 | 0.978 | 0.090 | 1.15 | 0.990 | 0.180 | 2.51 | 0.952 | 0.066 | 2.22 | 0.987 | 0.032 | 1.99 | 0.991 | 0.077 |
| EUC3 | 1.89 | 0.981 | 0.088 | 1.14 | 0.985 | 0.129 | 2.53 | 0.983 | 0.039 | 2.35 | 0.969 | 0.082 | 1.76 | 0.999 | 0.009 |
| EUC2 | 1.95 | 0.998 | 0.062 | 1.50 | 0.997 | 0.040 | 2.18 | 0.979 | 0.080 | 2.53 | 0.949 | 0.109 | 1.54 | 0.986 | 0.164 |
| EUC1 | 1.96 | 0.986 | 0.104 | 1.42 | 0.991 | 0.068 | 2.67 | 0.917 | 0.067 | 2.54 | 0.953 | 0.098 | 1.81 | 0.999 | 0.026 |

**Table 2**